\documentclass[english]{article}
\usepackage[T1]{fontenc}
\usepackage[latin9]{inputenc}
\usepackage{color}
\usepackage{amssymb}

\makeatletter
\newcommand{\lyxaddress}[1]{
\par {\raggedright #1
\vspace{1.4em}
\noindent\par}
}

\makeatother

\usepackage{babel}
\begin{document}

\title{\textbf{Precise model of Hawking radiation from the tunnelling mechanism }}

\author{\textbf{Christian Corda}}

\maketitle

\lyxaddress{\begin{center}
Dipartimento di Fisica, Scuola Superiore di Studi Universitari e
Ricerca \textquotedbl{}Santa Rita\textquotedbl{}, via Trasaghis 18/E,
00188 Roma, Italy
\par\end{center}}

\lyxaddress{\begin{center}
Austro-Ukrainian Institute for Science and Technology, Institut für
Theoretishe Physik, Technische Universität, Wiedner Hauptstrasse 8-10/136,
A-1040, Wien, Austria 
\par\end{center}}

\lyxaddress{\begin{center}
International Institute for Applicable Mathematics \& Information
Sciences (IIAMIS),  B.M. Birla Science Centre, Adarsh Nagar, Hyderabad
- 500 463, India
\par\end{center}}

\begin{center}
\textit{E-mail address:} \textcolor{blue}{cordac.galilei@gmail.com} 
\par\end{center}
\begin{abstract}
We recently improved the famous result of Parikh and Wilczek, who
found a probability of emission of Hawking radiation which is compatible
with a non-strictly thermal spectrum, showing that such a probability
of emission is really associated to two non-strictly thermal distributions
for boson and fermions. Here we finalize the model by finding the
correct value of the pre-factor of the Parikh and Wilczek probability
of emission. In fact, that expression has the $\sim$ sign instead
of the equality. In general, in this kind of leading order tunnelling
calculations, the exponent arises indeed from the classical action
and the pre-factor is an order Planck constant correction. But in
the case of emissions of Hawking quanta, the variation of the Bekenstein-Hawking
entropy is order 1 for an emitted particle having energy of order
the Hawking temperature. As a consequence, the exponent in the Parikh
and Wilczek probability of emission is order unity and one asks what
is the real significance of that scaling if the pre-factor is unknown.
Here we solve the problem assuming the unitarity of the black hole
(BH) quantum evaporation and considering the natural correspondence
between Hawking radiation and quasi-normal modes (QNMs) of excited
BHs , in a ``Bohr-like model'' that we recently discussed in a series
of papers. In that papers, QNMs are interpreted as natural BH quantum
levels (the ``electron states'' in the ``Bohr-like model''). Here
we find the intriguing result that, although in general it is well
approximated by 1, the pre-factor of the Parikh and Wilczek probability
of emission depends on the BH quantum level $n.$ We also write down
an elegant expression of the probability of emission in terms of the
BH quantum levels.
\end{abstract}
Hawking radiation \cite{key-1} is today studied in an elegant and
largely used way through the tunnelling mechanism, see \cite{key-2,key-3,key-4,key-5,key-6,key-7,key-8,key-41}
and refs. within. Let us see how that mechanism works. One considers
an object which is classically stable. If it becomes unstable from
a quantum-mechanically point of view, one naturally suspects tunnelling.
Hawking's famous mechanism of particles creation by BH \cite{key-1}
is, in turn, described in terms of tunnelling arising from vacuum
fluctuations near the BH horizon \cite{key-2,key-3,key-4,key-5,key-6,key-7,key-8,key-41}.
When a virtual particle pair is created just inside the BH horizon,
the virtual particle having positive energy can tunnel out. Thus,
it materializes outside the BH as a real particle. In analogous way,
when a virtual particle pair is created just outside the horizon,
the particle having negative energy can tunnel inwards. In both cases,
the particle having negative energy is absorbed by the BH. The result
is that the BH mass decreases and the particle having positive energy
propagates towards infinity. Thus, subsequent emissions of Hawking
quanta appear.

\noindent Working with $G=c=k_{B}=\hbar=\frac{1}{4\pi\epsilon_{0}}=1$
(Planck units), in strictly thermal approximation the probability
of emission of Hawking quanta is \cite{key-1,key-2,key-3,key-9} 
\begin{equation}
\Gamma\sim\exp(-\frac{\omega}{T_{H}}),\label{eq: hawking probability}
\end{equation}

\noindent where $\omega$ is the energy-frequency of the emitted particle
and $T_{H}\equiv\frac{1}{8\pi M}$ is the Hawking temperature. Taking
into account the energy conservation, i.e. the BH contraction enabling
a varying BH geometry, one gets the remarkable correction of Parikh
and Wilczek \cite{key-2,key-3} 
\begin{equation}
\Gamma\sim\exp[-\frac{\omega}{T_{H}}(1-\frac{\omega}{2M})]\quad\Longrightarrow\quad\Gamma=\alpha\exp[-\frac{\omega}{T_{H}}(1-\frac{\omega}{2M})],\label{eq: Parikh Correction}
\end{equation}
where $\alpha\sim1$ and the additional term $\frac{\omega}{2M}\:$
is present. We have recently improved the Parikh and Wilczek tunnelling
picture showing that the probability of emission (\ref{eq: Parikh Correction})
is, indeed, associated to the two distributions \cite{key-8} 
\begin{equation}
\begin{array}{c}
<N>_{boson}=\frac{1}{\exp\left[4\pi\left(2M-\omega\right)\omega\right]-1}\\
\\
<N>_{fermion}=\frac{1}{\exp\left[4\pi\left(2M-\omega\right)\omega\right]+1},
\end{array}\label{eq: final distributions}
\end{equation}
for bosons and fermions respectively, which are \emph{non} strictly
thermal. 

We note that eq. (\ref{eq: Parikh Correction}) has the $\sim$ sign
instead of the equality. In fact, in this kind of leading order tunnelling
calculations the exponent arises from the classical action and the
pre-factor is an order Planck constant correction. But in the case
of emissions of Hawking quanta the variation of the Bekenstein-Hawking
entropy \cite{key-2,key-3} 
\begin{equation}
\Gamma=\alpha\exp\Delta S_{BH}=\alpha\exp[-\frac{\omega}{T_{H}}(1-\frac{\omega}{2M})],\label{eq: variation BH entropy}
\end{equation}
is order 1 for an emitted particle having energy of order the Hawking
temperature. As a consequence, the exponent in the right hand side
of eqs. (\ref{eq: Parikh Correction}) and (\ref{eq: variation BH entropy})
is order unity and we ask what is the real significance of that scaling
if pre-factor is unknown. Here we solve the problem considering the
natural correspondence between Hawking radiation and BH QNMs, in a
``Bohr-like model'' that we recently discussed in a series of papers
\cite{key-9,key-10,key-11,key-12}, also together with collaborators
\cite{key-13,key-14}. 

We consider Dirac delta perturbations \cite{key-9,key-10,key-11,key-12},
which represent subsequent absorptions of particles having negative
energies. Such perturbations are associated to emissions of Hawking
quanta in the above discussed mechanism of particle pair creation.
BH response to perturbations are QNMs \cite{key-9,key-10,key-11,key-12,key-13,key-14,key-15,key-16},
which are frequencies of radial spin-$j$ perturbations obeying a
time independent Schröedinger-like equation \cite{key-9,key-10,key-11,key-12,key-16}.
They are the BH modes of energy dissipation which frequency is allowed
to be complex\cite{key-9,key-10,key-11,key-12,key-16}. The intriguing
idea to model the quantum BH in terms of BH QNMs arises from a remarkable
paper by York \cite{key-17}. For large values of the quantum ``overtone''
number $n$, where $n=1,2,...$, QNMs become independent of both the
spin and the angular momentum quantum numbers \cite{key-9,key-10,key-11,key-12,key-15,key-16},
in perfect agreement with \emph{Bohr's Correspondence Principle} \cite{key-18},
which states that \textquotedblleft{}transition frequencies at large
quantum numbers should equal classical oscillation frequencies\textquotedblright{}.
Thus, Bohr's Correspondence Principle enables an accurate semi-classical
analysis for large values of the principal quantum number $n,$ i.e,
for excited BHs. By using that principle, Hod has shown that QNMs
release information about the area quantization as QNMs are associated
to absorption of particles \cite{key-19,key-20}. Hod's work was refined
by Maggiore \cite{key-15} who solved some important problems. On
the other hand, as QNMs are \emph{countable} frequencies, ideas on
the \emph{continuous} character of Hawking radiation did not agree
with attempts to interpret QNMs in terms of emitted quanta, preventing
to associate QNMs to Hawking radiation \cite{key-16}. Recently, the
authors of \cite{key-21,key-22,key-23,key-24} and ourselves and collaborators
\cite{key-10,key-11,key-12,key-13,key-14} observed that the non-thermal
spectrum of Parikh and Wilczek \cite{key-2,key-3} also implies the
countable character of subsequent emissions of Hawking quanta. This
issue enables a natural correspondence between QNMs and Hawking radiation,
permitting to interpret QNMs also in terms of emitted energies \cite{key-10,key-11,key-12,key-13,key-14}.
Dirac delta perturbations due to discrete subsequent absorptions of
particles having negative energies, which are associated to emissions
of Hawking quanta in the mechanism of particle pair creation by quantum
fluctuations, generates indeed BH QNMs \cite{key-10,key-11,key-12,key-13,key-14}.
In other words, the BH contraction due to the energy conservation
is not a ``one shot process''. It generates oscillations of the
horizon instead, which are the QNMs. We also stress that the correspondence
between emitted radiation and proper oscillation of the emitting body
is a fundamental behavior of every radiation process in science. Based
on such a natural correspondence between Hawking radiation and BH
QNMs, one can consider QNMs in terms of quantum levels also for emitted
energies \cite{key-10,key-11,key-12,key-13,key-14}. 

Let us see how the model works. By introducing the \emph{effective
temperature }\cite{key-8,key-10,key-11,key-12} 
\begin{equation}
T_{E}(\omega)\equiv\frac{2M}{2M-\omega}T_{H}=\frac{1}{4\pi(2M-\omega)},\label{eq: Corda Temperature}
\end{equation}
one rewrites eq. (\ref{eq: Corda Temperature}) in a Boltzmann-like
form similar to eq. (\ref{eq: hawking probability}) 
\begin{equation}
\Gamma=\alpha\exp[-\beta_{E}(\omega)\omega]=\alpha\exp(-\frac{\omega}{T_{E}(\omega)}),\label{eq: Corda Probability}
\end{equation}

\noindent where $\exp[-\beta_{E}(\omega)\omega]$ is the \emph{effective
Boltzmann factor,} with $\beta_{E}(\omega)\equiv\frac{1}{T_{E}(\omega)}$.
Thus, the effective temperature replaces the Hawking temperature in
the equation of the probability of emission \cite{key-8,key-10,key-11,key-12}.
We emphasize that there are various fields of science where one takes
into account the deviation from the thermal spectrum of an emitting
body by introducing an effective temperature which represents the
temperature of a black body that would emit the same total amount
of radiation\emph{.} We introduced the concept of effective temperature
in BH physics in \cite{key-10,key-11} and used it in \cite{key-8,key-10,key-11,key-12}
and, together with collaborators, in \cite{key-13,key-14}. The effective
temperature depends on the energy-frequency of the emitted radiation
and the ratio $\frac{T_{E}(\omega)}{T_{H}}=\frac{2M}{2M-\omega}$
represents the deviation of the BH radiation spectrum from the strictly
thermal feature \cite{key-8,key-10,key-11,key-12}. The introduction
of the effective temperature permits the introduction of other \emph{effective
quantities}. Considering the initial BH mass \emph{before} the emission,
$M$, and the final BH mass \emph{after} the emission, $M-\omega$,
one introduces the \emph{BH} \emph{effective mass }and the \emph{BH
effective horizon} \cite{key-8,key-10,key-11,key-12} as 
\begin{equation}
M_{E}\equiv M-\frac{\omega}{2},\mbox{ }r_{E}\equiv2M_{E},\label{eq: effective quantities}
\end{equation}

\noindent \emph{during} the BH contraction, i.e. \emph{during} the
emission of the particle \cite{key-10}-\cite{key-12}. Such effective
quantities are average quantities \cite{key-8,key-10,key-11,key-12}.
In fact, \emph{$r_{E}$ }is the average of the initial and final horizons
while \emph{$M_{E}$ }is the average of the initial and final masses
\cite{key-8,key-10,key-11,key-12}. The effective temperature \emph{$T_{E}\:$
}is the inverse of the average value of the inverses of the initial
and final Hawking temperatures (\emph{before} the emission $T_{H\mbox{ initial}}=\frac{1}{8\pi M}$,
\emph{after} the emission $T_{H\mbox{ final}}=\frac{1}{8\pi(M-\omega)}$)
\cite{key-8,key-10,key-11,key-12}. 

For large values of the principal quantum number $n,$ i.e, for excited
BHs, and independently of the angular momentum quantum number, the
QNMs expression of the Schwarzschild BH which takes into account the
non-strictly thermal behavior of the radiation spectrum is obtained
replacing the Hawking temperature with the effective temperature in
the standard, strictly thermal, equation for the quasi-normal frequencies
as \cite{key-8,key-10,key-11,key-12}

\noindent 
\begin{equation}
\begin{array}{c}
\omega_{n}=a+ib+2\pi in\times T_{E}(|\omega_{n}|)\\
\\
\backsimeq2\pi in\times T_{E}(|\omega_{n}|)=\frac{in}{4M-2|\omega_{n}|},
\end{array}\label{eq: quasinormal modes corrected}
\end{equation}
where $a$ and $b$ are real numbers with $a=(\ln3)\times T_{E}(|\omega_{n}|),\; b=\pi\times T_{E}(|\omega_{n}|)$
for $j=0,2$ (scalar and gravitational perturbations), $a=0,\; b=0$
for $j=1$ (vector perturbations) and $a=0,\; b=\pi\times T_{E}(|\omega_{n}|)$
for half-integer values of $j$. On the other hand, as $a,b\ll|2\pi inT_{E}(|\omega_{n}|)|$,
a fundamental consequence is that the quantum of area obtained from
the asymptotics of $|\omega_{n}|$ is an intrinsic property of Schwarzschild
BHs because for large $n$ the leading asymptotic behavior of $|\omega_{n}|$
is given by the leading term in the imaginary part of the complex
frequencies, and it does not depend on the spin content of the perturbation
\cite{key-10,key-11,key-12,key-15}. 

\noindent An intuitive derivation of eq. (\ref{eq: quasinormal modes corrected})
can be found in \cite{key-10,key-11}. We \emph{rigorously} derived
such an equation in the Appendix of \cite{key-12}. Further important
clarifications on the derivation of eq. (\ref{eq: quasinormal modes corrected})
have been highlighted in the recent review paper \cite{key-37} through
Hawking's periodicity arguments \cite{key-38}.

\noindent Eq. (\ref{eq: quasinormal modes corrected}) has the following
elegant interpretation \cite{key-10,key-11}. The quasi-normal frequencies
determine the position of poles of a Green's function on the given
background, and the Euclidean BH solution converges to a \emph{non-strictly}
thermal circle at infinity with the inverse temperature $\beta_{E}(\omega_{n})=\frac{1}{T_{E}(|\omega_{n}|)}$
\cite{key-10,key-11}. Thus, the spacing of the poles in eq. (\ref{eq: quasinormal modes corrected})
coincides with the spacing $2\pi iT_{E}(|\omega_{n}|)=2\pi iT_{H}(\frac{2M}{2M-|\omega_{n}|}),$
expected for a \emph{non-strictly} thermal Green's function \cite{key-10,key-11}.
We found the physical solution for the absolute values of the frequencies
(\ref{eq: quasinormal modes corrected}) in \cite{key-10,key-11,key-12}.
Considering the leading asymptotic behavior one gets \cite{key-10,key-11,key-12}

\noindent 
\begin{equation}
E_{n}\equiv|\omega_{n}|=M-\sqrt{M^{2}-\frac{n}{2}}.\label{eq: radice fisica}
\end{equation}
$E_{n}\:$ is interpreted like the total energy emitted by the BH
at that time, i.e. when the BH is excited at a level $n$ \cite{key-10,key-11,key-12}.
Considering an emission from the ground state (i.e. a BH which is
not excited) to a state with large $n=n_{1}$ and using eq. (\ref{eq: radice fisica}),
the BH mass changes from $M\:$ to \cite{key-10,key-11,key-12}

\begin{equation}
M_{n_{1}}\equiv M-E_{n_{1}}=\sqrt{M^{2}-\frac{n_{1}}{2}}.\label{eq: me-1}
\end{equation}
In the transition from the state with $n=n_{1}$ to a state with $n=n_{2}$
where $n_{2}>n_{1}$ the BH mass changes again from $M_{n_{1}}\:$
to

\begin{equation}
\begin{array}{c}
M_{n_{2}}\equiv M_{n_{1}}-\Delta E_{n_{1}\rightarrow n_{2}}=M-E_{n_{2}}\\
=\sqrt{M^{2}-\frac{n_{2}}{2}},
\end{array}\label{eq: me}
\end{equation}
where 
\begin{equation}
\Delta E_{n_{1}\rightarrow n_{2}}\equiv E_{n_{2}}-E_{n_{1}}=M_{n_{1}}-M_{n_{2}}=\sqrt{M^{2}-\frac{n_{1}}{2}}-\sqrt{M^{2}-\frac{n_{2}}{2}},\label{eq: jump}
\end{equation}
is the jump between the two levels due to the emission of a particle
having frequency $\Delta E_{n_{1}\rightarrow n_{2}}$. Thus, in our
BH model \cite{key-12}, during a quantum jump a discrete amount of
energy is radiated and, for large values of the principal quantum
number $n,$ the analysis becomes independent of the other quantum
numbers. In a certain sense, QNMs represent the \textquotedbl{}electron\textquotedbl{}
which jumps from a level to another one and the absolute values of
the QNMs frequencies represent the energy \textquotedbl{}shells\textquotedbl{}
{[}2{]}. In Bohr model of the hydrogen atom {[}25, 26{]} electrons
can only gain and lose energy by jumping from one allowed energy shell
to another, absorbing or emitting radiation with an energy difference
of the levels according to the Planck relation (in standard units)
$E=hf$, where $\: h\:$ is the Planck constant and $f\:$ the transition
frequency. In our Bohr-like BH model \cite{key-10,key-11,key-12},
QNMs can only gain and lose energy by jumping from one allowed energy
shell to another, absorbing or emitting radiation (emitted radiation
is given by Hawking quanta) with an energy difference of the levels
according to eq. (\ref{eq: jump}). The similarity is completed if
one notes that the interpretation of eq. (\ref{eq: radice fisica})
is of a particle, the ``electron'', quantized on a circle of length
\cite{key-10,key-11} 
\begin{equation}
L=\frac{1}{T_{E}(E_{n})}=4\pi\left(M+\sqrt{M^{2}-\frac{n}{2}}\right),\label{eq: lunghezza cerchio}
\end{equation}
which is the analogous of the electron travelling in circular orbits
around the hydrogen nucleus, similar in structure to the solar system,
of Bohr model {[}25, 26{]}. On the other hand, Bohr model is an approximated
model of the hydrogen atom with respect to the valence shell atom
model of full quantum mechanics. In the same way, our Bohr-like BH
model {[}12{]} should be an approximated model with respect to the
definitive, but at the present time unknown, BH model arising from
a full quantum gravity theory. 

\noindent As $E_{n}$ is interpreted like the total energy emitted
at level $n$ {[}12{]}, considering the expressions (\ref{eq: me-1})
and (\ref{eq: me}) for the residual BH mass one needs also {[}12{]}

\begin{equation}
M^{2}-\frac{n}{2}\geq0.\label{eq: need}
\end{equation}
In fact, BHs cannot emit more energy than their total mass and the
total energy emitted by the BH cannot be imaginary. The expression
(\ref{eq: need}) gives a maximum value for the overtone number $n$ 

\begin{equation}
n\leq n_{max}=2M^{2},\label{eq: n max}
\end{equation}
which corresponds to $E_{n_{max}}=M.$ On the other hand, we recall
that, by using the Generalized Uncertainty Principle, Adler, Chen
and Santiago {[}27{]} have shown that the total BH evaporation is
prevented in exactly the same way that the Uncertainty Principle prevents
the hydrogen atom from total collapse. In fact, the collapse is prevented,
not by symmetry, but by dynamics, as the \emph{Planck distance} and
the \emph{Planck mass} are approached {[}27{]}. That important result
implies that eq. (\ref{eq: need}) has to be slightly modified, becoming
(the \emph{Planck mass} is equal to $1$ in Planck units) {[}12{]}

\begin{equation}
M^{2}-\frac{n}{2}\geq1.\label{eq: need 1}
\end{equation}
Thus, one gets a slightly different value of the maximum value of
the overtone number $n$ 

\begin{equation}
n\leq n_{max}=2(M^{2}-1).\label{eq: n max 1}
\end{equation}
Then, the countable sequence of QNMs for emitted energies cannot be
infinity although $n$ can be extremely large {[}12{]}. In fact, restoring
ordinary units and considering a BH mass of the order of 10 solar
masses, one easily gets $n_{max}\sim10^{76}.$ On the other hand,
we expect further corrections to our semi-classical analysis when
the \emph{Planck scale} is approached, as we need a full theory of
quantum gravity to obtain a correct description of the Planck scale's
physics.

Our Bohr-like BH model in \cite{key-10,key-11,key-12} is in full
agreement with previous literature of BH thermodynamics, like references
\cite{key-15,key-28,key-29}. More, it is also in full agreement with
the famous result of Bekenstein on the \emph{area quantization} \cite{key-30}.
In fact, we found an area quantum arising from a jump among two neighbouring
quantum levels $n-1$ and $n$ having a value $|\triangle A_{n}|=|\triangle A_{n-1}|\simeq8\pi,$
see eq. (37) in \cite{key-12}, which is totally consistent with Bekenstein's
result \cite{key-30}. Clearly, all these similarities with the Bohr
semi-classical model of the hydrogen atom and all these consistences
with well known results in the literature of BHs, starting by the
universal Bekenstein's result, \emph{cannot} be coincidences, but
are confirmations of the correctness of the analysis in \cite{key-10,key-11,key-12}
instead. 

Concerning the important issue of the BH entropy, we recall the recent
interesting result in which the entropy is connected with the tunnelling
mechanism \cite{key-40,key-42}.  Jiang and Han \cite{key-40} have
indeed quantized the BH entropy by combining the proposal about the
BH adiabatic invariance and the proposal about the oscillating velocity
of the BH horizon, where the velocity is obtained in the tunneling
framework.

Now, let us proceed in calculating the correct value of the pre-factor
of eqs. (\ref{eq: Parikh Correction}) and (\ref{eq: variation BH entropy}).
We recall that, today, the majority of researchers thinks that BH
quantum evaporation is an unitary process and that Hawking's original
claim on the information loss in BH evaporation \cite{key-9} was
wrong. Various approaches are indeed proposed by various researchers
in order to solve the BH information paradox and to recover unitarity
in BH evaporation. Here we recall: i) the approach of \cite{key-21,key-22,key-23,key-24}
where the authors found the existence of correlations among Hawking
radiation which are elegantly described as hidden messengers in BH
evaporation permitting to restore unitarity in gravitational collapse;
ii) the famous ADS/CFT correspondence \cite{key-31}, which was endorsed
by both Susskind \cite{key-32} and Hawking \cite{key-33}, who reversed
his opinion in 2004 and recently claimed that BH evaporation is unitary
\cite{key-34}; iii) the approach of the so called ``fuzzballs''
\cite{key-35}; iv) our recent approach based on the time evolution
of our Bohr-like BH model \cite{key-39}. We have indeed shown in
\cite{key-39} that the time evolution of our Bohr-like BH model is
governed by a \emph{time dependent Schrödinger equation }for the system
composed by Hawking radiation and BH QNMs. The physical state and
the correspondent\emph{ wave function }are written in terms of an
\emph{unitary} evolution matrix instead of a density matrix \cite{key-39}.
In that way, the final state results to be a \emph{pure} quantum state
instead of a mixed one \cite{key-39}. The approach in \cite{key-39}
permits also to solve the entanglement problem connected with the
information paradox because emitted Hawking quanta results to be entangled
with BH QNMs. Thus, hereafter we will assume the unitarity of BH quantum
evaporation. Following \cite{key-39}, now we show that, fixed two
quantum levels $m$ and $n$, the energy emitted in an arbitrary transition
$m\rightarrow n$, with $n>m$, is proportional to the effective temperature
associated to the transition and that the constant of proportionality
depends only on the difference $m-n$. Setting \cite{key-39} 
\begin{equation}
\Delta E_{m\rightarrow n}\equiv E_{n}-E_{m}=M_{m}-M_{n}=K\left[T_{E}\right]_{m\rightarrow n},\label{eq: differenza radici fisiche}
\end{equation}
where $M_{m}$ and $M_{n}$ are given by eqs. (\ref{eq: me-1}) and
(\ref{eq: me}), let us see if there are values of the constant $K$
for which eq. (\ref{eq: differenza radici fisiche}) is satisfied.
We recall that \cite{key-39} 

\begin{equation}
\left[T_{E}\right]_{m\rightarrow n}=\frac{1}{4\pi\left(M_{m}+M_{n}\right)},\label{eq: temperatura efficace di transizione}
\end{equation}
because the effective temperature is the inverse of the average value
of the inverses of the initial and final Hawking temperatures, see
the above discussion . Thus, eq. (\ref{eq: differenza radici fisiche})
can be rewritten as \cite{key-39} 

\begin{equation}
M_{m}^{2}-M_{n}^{2}=\frac{K}{4\pi}.\label{eq: differenza radici fisiche 2}
\end{equation}
By using eqs. (\ref{eq: me-1}) and (\ref{eq: me}), eq. (\ref{eq: differenza radici fisiche 2})
becomes \cite{key-39} 

\begin{equation}
\frac{1}{2}\left(n-m\right)=\frac{K}{4\pi},\label{eq: K solved}
\end{equation}
which implies that eq. (\ref{eq: differenza radici fisiche}) is satisfied
for $K=2\pi\left(n-m\right).$ Hence, one finds \cite{key-39} 
\begin{equation}
\Delta E_{m\rightarrow n}=E_{n}-E_{m}=2\pi\left(n-m\right)\left[T_{E}(\omega)\right]_{m\rightarrow n}.\label{eq: differenza radici fisiche finale}
\end{equation}
Using eq. (\ref{eq: Corda Probability}), the probability of emission
between the two levels $n$ and $m$ can be written in the intriguing
form \cite{key-39} 
\begin{equation}
\Gamma_{m\rightarrow n}=\alpha\exp-\left\{ \frac{\Delta E_{m\rightarrow n}}{\left[T_{E}(\omega)\right]_{m\rightarrow n}}\right\} =\alpha\exp\left[-2\pi\left(n-m\right)\right].\label{eq: Corda Probability Intriguing}
\end{equation}
Thus, the probability of emission between two arbitrary levels characterized
by the two ``overtone'' quantum numbers $m$ and $n$ scales like
$\exp\left[-2\pi\left(n-m\right)\right].$ In particular, for $n=m+1$
the probability of emission has its maximum value $\sim\exp(-2\pi)$,
i.e. the probability is maximum for two adjacent levels, as one can
intuitively expect \cite{key-39} . If one fixes $m$, the assumption
of unitarity in BH evaporation permits the probabilities (\ref{eq: Corda Probability Intriguing})
to be normalized to the unity as

\begin{equation}
\sum_{n=m}^{n_{max}}\Gamma_{m\rightarrow n}=\sum_{n=m}^{n_{max}}\alpha\exp\left[-2\pi\left(n-m\right)\right]=1,\label{eq: normalizzazione}
\end{equation}
where $n_{max}$ is the maximum value for the ``overtone'' number
$n$ given by eq. (\ref{eq: n max 1}) and $n=m$ corresponds to the
probability that the BH does not emit. Putting $k=n-m\:$ and $\exp\left[-2\pi\right]=X\:$
eq. (\ref{eq: normalizzazione}) becomes

\begin{equation}
\sum_{k=0}^{k_{max}}\Gamma_{0\rightarrow k}=\alpha\sum_{k=0}^{k_{max}}X^{k}=1.\label{eq: normalizzazione 2}
\end{equation}
The sum in eq. (\ref{eq: normalizzazione 2}) is the $kth$ partial
sum of the geometric series and can be solved as \cite{key-36}

\begin{equation}
\sum_{k=0}^{k_{max}}X^{k}=\frac{1-X^{\left(k_{max}+1\right)}}{1-X}.\label{eq: serie geometrica}
\end{equation}
Thus, one gets 
\begin{equation}
\alpha\frac{1-X^{\left(k_{max}+1\right)}}{1-X}=1,\label{eq: serie geometrica  risolta}
\end{equation}
which permits to solve for $\alpha$ 

\begin{equation}
\alpha\equiv\alpha_{m}=\frac{1-X}{1-X^{\left(k_{max}+1\right)}}=\frac{1-\exp\left[-2\pi\right]}{1-\exp\left[-2\pi\left(n_{max}-m+1\right)\right]}.\label{eq: mio alpha number}
\end{equation}
Hence, we find that the pre-factor $\alpha$ depends on the BH quantum
level $m.$ Inserting the result (\ref{eq: mio alpha number}) in
eq. (\ref{eq: Corda Probability Intriguing}) we fix the probability
of emission between the two levels $m$ and $n$ as

\begin{equation}
\begin{array}{c}
\Gamma_{m\rightarrow n}=\alpha_{m}\exp-\left\{ \frac{\Delta E_{m\rightarrow n}}{\left[T_{E}(\omega)\right]_{m\rightarrow n}}\right\} =\alpha_{m}\exp\left[-2\pi\left(n-m\right)\right]=\\
\\
=\left\{ \frac{1-\exp\left[-2\pi\right]}{1-\exp\left[-2\pi\left(n_{max}-m+1\right)\right]}\right\} \exp\left[-2\pi\left(n-m\right)\right].
\end{array}\label{eq: Corda Probability Intriguing finalized}
\end{equation}
From the quantum mechanical point of view, one can physically interpret
Hawking radiation like energies of quantum jumps among the unperturbed
levels (\ref{eq: radice fisica}) \cite{key-8,key-10,key-11,key-12,key-13}. 

It might be a benefit for the reader to rewrite eq. (30) in terms
of $\omega$ and $M$ using $n_{max}=2M^{2}$ and $2M^{2}-m=2(M-\omega_{m})^{2}$.
One gets:

\begin{equation}
\Gamma_{m\rightarrow n}=\left\{ \frac{1-\exp\left[-2\pi\right]}{1-\exp\left[-4\pi(M-\omega_{m})^{2}+2\pi\right]}\right\} \exp\left[-2\pi\left(n-m\right)\right].\label{eq: benefit}
\end{equation}
Eq. (31) can be useful when exploring $n=m+1$ expressions throughout
the spectrum as a function of $\omega_{m}/2M$ \emph{.}

In any case, we notice from eq. (\ref{eq: mio alpha number}) that
\begin{equation}
\alpha_{m}\simeq1-\exp\left[-2\pi\right]\simeq1\quad for\quad n_{max}\gg m\label{eq: for}
\end{equation}
and, for increasing $m$ 
\begin{equation}
\alpha_{m}\rightarrow1^{-}\quad for\quad m\rightarrow n_{max}^{-}\label{eq: for 2}
\end{equation}
with 
\begin{equation}
\alpha_{m}=1\quad for\quad m=n_{max}.\label{eq: for 3}
\end{equation}
In other words, the dependence of the pre-factor $\alpha$ on the
BH quantum level $m$ is in general well approximated by 1 in all
the process of BH evaporation for which our analysis works, included
late times in the evaporation process. Clearly, as we are using a
semi-classical approximation, deviations could be present at the Planck
scale. On the other hand, we need a full theory of quantum gravity
in order to achieve the Planck scale physics.

The analysis in this work is strictly correct only for $n\gg1,$ i.e.
only for excited BHs. This is the reason because we assumed an emission
from the ground state to a state with large $n\:$ in the discussion.
On the other hand, a state with large $n\:$ is always reached at
late times, maybe not through a sole emission from the ground state,
but, indeed, through various subsequent emissions of Hawking quanta.
For the sake of completeness, it is better to explicitly describe
in which part of this process needed to reach a large $n$ the description
in this paper would start to be valid \cite{key-42}. Let us consider
again an astrophysics BH having an original mass $M$ of the order
of 10 solar masses. By inserting $n=10^{6}$ in eq. (\ref{eq: radice fisica})
one gets $E_{n}\simeq10^{-27}M_{Planck}$. In other words, the BH
lost a negligible part of his mass. Thus, we understand that our description
is valid for almost all the process of BH evaporation. 

It is also interesting to include a comparison between the importance
of the deviation from thermality described in this paper and that
due to other phenomena, such as the backscattering of Hawking radiation
on the Schwarzschild metric (which is already present at early times
in the evaporation) \cite{key-42}. This is a different physical phenomenon
due to the existence of a BH potential barrier \cite{key-43,key-44}.
Let us consider a scalar field $\Phi$ in the Schwarzschild space-time
\cite{key-43}. As this space-time is spherically symmetric, one can
separate the Klein-Gordon equation governing the scalar field, i.e.
\cite{key-43}

\begin{equation}
\left(\square+m^{2}\right)\Phi=0,\label{eq: Klein-Gordon}
\end{equation}
into spherical harmonics \cite{key-43} 
\begin{equation}
\Phi=\frac{f\left(r,t\right)}{r}Y_{lm}\exp\left[-i\omega t\right].\label{eq: spherical harmonics}
\end{equation}
 Introducing the Regge-Wheeler tortoise coordinate $x$, defined through
the relation \cite{key-37} 
\begin{equation}
\begin{array}{c}
x=r+2M\ln\left(\frac{r}{2M}-1\right)\\
\\
\frac{\partial}{\partial x}=\left(1-\frac{2M}{r}\right)\frac{\partial}{\partial r},
\end{array}\label{eq:original  tortoise}
\end{equation}
one gets the resulting radial wave equation as \cite{key-43} 
\begin{equation}
\frac{\partial^{2}f}{\partial t^{2}}-\frac{\partial^{2}f}{\partial x^{2}}=V_{l}(r)f,\label{eq: radial wave equation}
\end{equation}
where the potential is \cite{key-43} 
\begin{equation}
V_{l}(r)\equiv\left(1-\frac{2M}{r}\right)\left[\frac{2M}{r^{3}}+\frac{l\left(l+1\right)}{r^{2}}+m^{2}\right],\label{eq: potenziale}
\end{equation}
and $m$ is the mass of the scalar field \cite{key-43}. Near the
BH horizon we get $x\rightarrow-\infty$. As a consequence, the potential
falls off exponentially \cite{key-43} 
\begin{equation}
V\sim\exp\left[\frac{x}{2M}\right].\label{eq: potenziale approssimato}
\end{equation}
On the other hand, for $x\rightarrow+\infty$ one gets \cite{key-43}
\begin{equation}
\begin{array}{c}
V\sim m^{2}\left(1-\frac{2M}{x}\right)\quad for\, massive\,\Phi\\
\\
V\sim\frac{l\left(l+1\right)}{r^{2}}\quad for\, massless\,\Phi.
\end{array}\label{eq: potenziale doppio}
\end{equation}
The gravitational field will be partially scattered off on the potential
(\ref{eq: potenziale}) by the incoming waves. Thus, we obtain a superposition
of incoming and outgoing waves \cite{key-43}. As a consequence, the
spectrum is not precisely thermal \cite{key-43}. 

Although the two phenomena generating deviation from the strict thermal
behavior of the Hawking radiation spectrum are different, we recall
an interesting recent work where both of them are taken into account
\cite{key-45}.

\subsection*{Conclusion remarks}

Assuming the unitarity of BH quantum evaporation \cite{key-21,key-22,key-23,key-24,key-31,key-32,key-33,key-34,key-35}
and considering the natural correspondence between Hawking radiation
and BH QNMs, in a ``Bohr-like model'' for excited BHs that we recently
discussed in a series of papers \cite{key-10,key-11,key-12,key-13},
we have found the intriguing result that the pre-factor of the Parikh
and Wilczek probability of emission, although if in general well approximated
by 1, depends on the BH quantum level $n.$ Then, one gets that the
emission of Hawking radiation, in the tunneling framework, is completely
determined by eqs. (\ref{eq: final distributions}) and (\ref{eq: Corda Probability Intriguing finalized}).

\subsection*{Acknowledgements}

The author thanks the Referees for useful comments which permitted
to improve this paper.


\begin{thebibliography}{37 }
\bibitem{key-1}S. W. Hawking, Commun. Math. Phys. 43, 199 (1975).

\bibitem[2]{key-2}M. K. Parikh and F. Wilczek, Phys. Rev. Lett. 85,
5042 (2000). 

\bibitem[3]{key-3}M. K. Parikh, Gen. Rel. Grav. 36, 2419 (2004).

\bibitem[4]{key-4}R. Banerjee and B.R. Majhi, JHEP 0806, 095 (2008).

\bibitem[5]{key-5}M. Angheben, M. Nadalini, L. Vanzo and S. Zerbini,
JHEP 0505, 014 (2005).

\bibitem[6]{key-6}M. Arzano, A. J. M. Medved and E. C. Vagenas, JHEP
0509, 037 (2005).

\bibitem[7]{key-7}R. Banerjee and B.R. Majhi, Phys. Lett. B 675,
243 (2009).

\bibitem[8]{key-8}C. Corda, Ann. Phys. 337, 49 (2013), final version
with corrected typos in arXiv:1305.4529v3.

\bibitem[9]{key-9}S. W. Hawking, Phys. Rev. D 14, 2460 (1976). 

\bibitem[10]{key-10}C. Corda, Int. Journ. Mod. Phys. D 21, 1242023
(2012).

\bibitem[11]{key-11}C. Corda, JHEP 1108, 101 (2011). 

\bibitem[12]{key-12}C. Corda, Eur. Phys. J. C 73, 2665 (2013).

\bibitem[13]{key-13}C. Corda, S. H. Hendi, R. Katebi, N. O. Schmidt,
06, 008 (2013).

\bibitem[14]{key-14}C. Corda, S. H. Hendi, R. Katebi, N. O. Schmidt,
Adv. High En. Phys. 527874 (2014).

\bibitem[15]{key-15}M. Maggiore, Phys. Rev. Lett. 100, 141301 (2008).

\bibitem[16]{key-16}L. Motl, Adv. Theor. Math. Phys. 6, 1135 (2003).

\bibitem[17]{key-17}J. York Jr., Phys. Rev. D28, 2929 (1983). 

\bibitem[18]{key-18}N. Bohr, Zeits. Phys. 2, 423 (1920). 

\bibitem[19]{key-19}S. Hod, Gen. Rel. Grav. 31, 1639 (1999).

\bibitem[20]{key-20}S. Hod, Phys. Rev. Lett. 81 4293 (1998).

\bibitem[21]{key-21}B. Zhang, Q.-Y. Cai, L. You,, and M. S. Zhan,
Phys. Lett. B 675, 98 (2009). 

\bibitem[22]{key-22}B. Zhang, Q.-Y. Cai, M. S. Zhan, and L. You,
Ann. Phys. 326, 350 (2011). 

\bibitem[23]{key-23}X.-K. Guo, Q.-Y., Cai, Int. Journ. Theor. Phys.,
published online, DOI 10.1007/s10773-014-2095-8 (2014). 

\bibitem[24]{key-24}B. Zhang, Q.-Y. Cai, M. S. Zhan, and L. You,
Int. Journ. Mod. Phys. D 22, 1341014 (2013).

\bibitem[25]{key-25}N. Bohr, Philos. Mag. 26 , 1 (1913). 

\bibitem[26]{key-26}N. Bohr, Philos. Mag. 26 , 476 (1913).

\bibitem[27]{key-27}R. J. Adler, P. Chen and D. I. Santiago, Gen.
Rel. Grav. 3, 2101 (2001). 

\bibitem[28]{key-28}S. Shankaranarayanan, Mod. Phys. Lett. A 23,
1975-1980 (2008). 

\bibitem[29]{key-29}J. Zhang, Phys. Lett. B 668, 353-356 (2008). 

\bibitem[30]{key-30}J. D. Bekenstein, Lett. Nuovo Cim. 11, 467 (1974). 

\bibitem[31]{key-31}J. M. Maldacena, Adv. Theor. Math. Phys 2, 231
(1998). 

\bibitem[32]{key-32}L. Susskind, \emph{The Black Hole War: My Battle
with Stephen Hawking to Make the World Safe for Quantum Mechanics},
Little, Brown and Company (2008).

\bibitem[33]{key-33}S. W. Hawking, Phys. Rev. D 72, 084013 (2005).

\bibitem[34]{key-34}S. W. Hawking, arXiv:1401.5761 (2014).

\bibitem[35]{key-35}Samir D. Mathur, arXiv:1108.0302v2 (hep-th). 

\bibitem[36]{key-36}B. Datta and A. N. Singh, Ind. Jour. Hist. Sci.
28, 103 (1993).

\bibitem[37 ]{key-37}C. Corda, Invited Review for the Advances in
High Energy Physics Special Issue ``Dark Atoms and Dark Radiation'',
edited by Maxim Khlopov, Konstantin Belotsky, Jean-René Cudell and
Chris Kouvaris (2015), pre-print in arXiv:1503.00565. 

\bibitem[38]{key-38}S. W. Hawking, \textquotedblleft{}The Path Integral
Approach to Quantum Gravity\textquotedblright{}, in General Relativity:
An Einstein Centenary Survey, eds. S. W. Hawking and W. Israel, (Cambridge
University Press, 1979). 

\bibitem[39]{key-39}C. Corda, Ann. Phys. 353, 71 (2015).

\bibitem[40]{key-40}Q. Q. Jiang and Y. Han, Phys.Lett. B 718 584
(2012). 

\bibitem[41]{key-41}S. Z. Yang, H. L. Li, Q. Q. Jiang and M. Q. Liu,
Sci China-Phys. Mech. Astron. 50, 2 249 (2007).

\bibitem[42]{key-42}Private communication with the referees.

\bibitem[43]{key-43}D. Deeg, ``\emph{Quantum Aspects of Black Holes}'',
edoc.ub.uni-muenchen.de/.../1/Deeg\_Dorothea.pdf.

\bibitem[44]{key-44}V. Mukhanov, A. Wipf, A. Zelnikov, Phys. Lett.
B 332 283 (1994).

\bibitem[45]{key-45}R. Torres, F. Fayos, O. Lorente-Espin, Phys.
Lett. B 720, 198 (2013).\end{thebibliography}
\end{document}